\documentclass[copyright,creativecommons,noderivs]{eptcs}
\providecommand{\event}{ML 2014}
\usepackage{breakurl}
\usepackage{amsmath}

\usepackage[usenames,dvipsnames]{xcolor}

\usepackage{color}
\definecolor{mauve}{rgb}{0.58,0,0.82}
\definecolor{maroon}{rgb}{0.5,0,0}

\usepackage{listings}
\lstdefinelanguage{scala}{
  morekeywords={abstract,case,catch,class,def,%
    do,else,extends,false,final,finally,%
    for,if,implicit,import,match,mixin,%
    new,null,object,override,package,%
    private,protected,requires,return,sealed,%
    super,this,throw,trait,true,try,%
    type,val,var,while,with,yield},
  otherkeywords={=>,<-,<\%,<:,>:,\#,@},
  sensitive=true,
  morecomment=[l]{//},
  morecomment=[n]{/*}{*/},
  morestring=[b]",
  morestring=[b]',
  morestring=[b]"""
}

\lstdefinelanguage{ocamlgrammar}{
  keywords={in,let,module,implicit,open,:},
  otherkeywords={:,=,\{,\},->},
  sensitive=true,
  alsoletter={-},
  keywordstyle=\color{blue}, 
  identifierstyle=\color{maroon}\rmfamily\textit,  
  literate={DEF}{::=\;\;}1 {LSTAR}{(}1 {RSTAR}{)$^*$}1
                {LOPT}{(}1 {ROPT}{)$^?$}1
}

\lstdefinestyle{ocamlgrammar}
{
  language=ocamlgrammar,
  basicstyle=\ttfamily,
  columns=flexible,
  mathescape=true,
}
\lstdefinestyle{ocaml}
{
  language=[Objective]Caml,
  escapeinside={(**}{*)},
  morekeywords={implicit},
  basicstyle=\ttfamily,
  commentstyle=\color{green},
  keywordstyle=\color{blue},
  stringstyle=\color{mauve},
}
\lstdefinestyle{haskell}
{
  language=Haskell,
  deletekeywords={show,Show,String,Int,Ord,Bool,insert,delete,union},
  otherkeywords={},
  basicstyle=\ttfamily,
  commentstyle=\color{green},
  keywordstyle=\color{orange},
  stringstyle=\color{mauve},
}
\lstdefinestyle{scala}
{
  language=Scala,
  basicstyle=\ttfamily,
  commentstyle=\color{green},
  keywordstyle=\color{WildStrawberry},
  stringstyle=\color{mauve},
}
\lstdefinestyle{cpp}
{
  language=C++,
  morekeywords={concept,concept_map},
}

\newcommand{\code}[2][ocaml]{\lstinline[style={#1}]{#2}}

\title{Modular implicits}
\author{Leo White \and Fr\'{e}d\'{e}ric Bour \and Jeremy Yallop}
\def\titlerunning{Modular implicits}
\def\authorrunning{Leo White, Fr\'{e}d\'{e}ric Bour \& Jeremy Yallop}

\begin{document}

\maketitle

\begin{abstract}
  We present \textit{modular implicits}, an extension to the OCaml
  language for ad-hoc polymorphism inspired by Scala implicits and
  modular type classes.
  Modular implicits are based on type-directed implicit module
  parameters, and elaborate straightforwardly into OCaml's first-class
  functors.
  Basing the design on OCaml's modules leads to a system that
  naturally supports many features from other languages with
  systematic ad-hoc overloading, including inheritance, instance
  constraints, constructor classes and associated types.
\end{abstract}

\section{Introduction}

A common criticism of OCaml is its lack of support for \emph{ad-hoc
  polymorphism}. The classic example of this is OCaml's separate
addition operators for integers (\code{+}) and floating-point numbers
(\code{+.}). Another example is the need for type-specific printing
functions (\code{print\_int}, \code{print\_string}, etc.) rather than a
single \code{print} function which works across multiple types.

In this paper, we propose a system for ad-hoc polymorphism in OCaml
based on using modules as type-directed implicit parameters. We describe
the design of this system, and compare it to systems for ad-hoc
polymorphism in other languages.

A prototype implementation of our proposal based on OCaml 4.02.0 has
been created and is available through the OPAM package
manager (Section~\ref{implementation}).

\subsection{Type classes and implicits}
\label{intro:type-classes-and-implicits}

Ad-hoc polymorphism allows the dynamic semantics of a program to be
affected by the types of values in that program. A program may have more
than one valid typing derivation, and which one is derived when
type-checking a program is an implementation detail of the
type-checker. Jones et al.~\cite{jones1997type} describe the following
important property:
\begin{quote}
  Every different valid typing derivation for a program leads to a
  resulting program that has the same dynamic semantics.
\end{quote}
This property is called \emph{coherence} and is a fundamental property
that must hold in a system for ad-hoc polymorphism.

\subsubsection{Type classes}
\label{intro:type-classes-and-implicits:type-classes}

Type classes in Haskell~\cite{DBLP:conf/popl/WadlerB89} have proved an
effective mechanism for supporting ad-hoc polymorphism. Type classes
provide a form of constrained polymorphism, allowing constraints to be
placed on type variables. For example, the \code[haskell]{show} function
has the following type:
\begin{lstlisting}[style=haskell]
    show :: Show a => a -> String
\end{lstlisting}
This indicates that the type variable \code[haskell]{a} can only be
instantiated with types which obey the constraint \code[haskell]{Show
  a}. These constraints are called \emph{type classes}. The
\code[haskell]{Show} type class is defined as\footnote{Some methods of
  \code[haskell]{Show} have been omitted for simplicity.}:
\begin{lstlisting}[style=haskell]
    class Show a where
      show :: a -> String
\end{lstlisting}
which specifies a list of methods which must be provided in order for a
type to meet the \code[haskell]{Show} constraint. The method
implementations for a particular type are specified by defining an
\emph{instance} of the type class. For example, the instance of
\code[haskell]{Show} for \code[haskell]{Int} is defined as:
\begin{lstlisting}[style=haskell]
    instance Show Int where
        show = showSignedInt
\end{lstlisting}

Constraints on a function's type can be inferred based on the use of
other constrained functions in the function's definition. For example,
if a \code[haskell]{show_twice} function uses the \code[haskell]{show}
function:
\begin{lstlisting}[style=haskell]
    show_twice x = show x ++ show x
\end{lstlisting}
then Haskell will infer that \code[haskell]{show_twice} has type
\code[haskell]{Show a => a -> String}.

Haskell's coherence in the presence of inferred type constraints
relies on type class instances being \emph{canonical} -- the program
contains at most one instance of a type class for each type. For
example, a Haskell program can only contain at most one instance of
\code[haskell]{Show Int}, and attempting to define two such instances
will result in a compiler error. Section~\ref{canonicity:abstraction}
describes why this property cannot hold in OCaml.

Type classes are implemented using a type-directed implicit
parameter-passing mechanism. Each constraint on a type is treated as a
parameter containing a dictionary of the methods of the type
class. The corresponding argument is implicitly inserted by the
compiler using the appropriate type class instance.

\subsubsection{Implicits}
\label{intro:type-classes-and-implicits:implicits}

Implicits in Scala~\cite{DBLP:conf/oopsla/OliveiraMO10} provide similar
capabilities to type classes via direct support for type-directed
implicit parameter passing. Parameters can be marked \code[scala]{implicit}
which then allows them to be omitted from function calls. For example, a
\code[scala]{show} function could be specified as:
\begin{lstlisting}[style=scala]
    def show[T](x : T)(implicit s : Showable[T]): String
\end{lstlisting}
where \code[scala]{Showable[T]} is a normal Scala type defined as:
\begin{lstlisting}[style=scala]
    trait Showable[T] { def show(x: T): String }
\end{lstlisting}
The \code[scala]{show} function can be called just like any other:
\begin{lstlisting}[style=scala]
    object IntShowable extends Showable[Int] {
      def show(x: Int) = x.toString
    }

    show(7)(IntShowable)
\end{lstlisting}
However, the second argument can also be elided, in which case its value
is selected from the set of definitions in scope which have been marked
\code[scala]{implicit}. For example, if the definition of
\code[scala]{IntShowable} were marked \code[scala]{implicit}:
\begin{lstlisting}[style=scala]
    implicit object IntShowable extends Showable[Int] {
      def show(x: Int) = x.toString
    }
\end{lstlisting}
then \code[scala]{show} can be called on integers without specifying the
second argument -- which will automatically be inserted as
\code[scala]{IntShowable} because it has the required type
\code[scala]{Showable[Int]}:
\begin{lstlisting}[style=scala]
    show(7)
\end{lstlisting}

Unlike constraints in Haskell, Scala's implicit parameters must always
be added to a function explicitly. The need for a function to have an
implicit parameter cannot be inferred from the function's
definition. Without such inference, Scala's coherence can rely on the
weaker property of \emph{non-ambiguity} instead of
\emph{canonicity}. This means that you can define multiple implicit
objects of type \code[scala]{Showable[Int]} in your program without
causing an error. Instead, Scala issues an error if the resolution of an
implicit parameter is ambiguous. For example, if two implicit objects of
type \code[scala]{Showable[Int]} are in scope when \code[scala]{show} is
applied to an \code[scala]{Int} then the compiler will report an
ambiguity error.

\subsubsection{Modular type classes}
\label{intro:type-classes-and-implicits:modular-type-classes}

Dreyer et al.~\cite{DBLP:conf/popl/DreyerHCK07} describe \emph{modular
  type classes}, a type class system which uses ML module types as type
classes and ML modules as type class instances.

As with traditional type classes, type class constraints on a function can
be inferred from the function's definition. Unlike traditional type
classes, modular type classes cannot ensure that type class instances
are canonical (see Section~\ref{canonicity:abstraction}). Maintaining
coherence in the presence of constraint inference without canonicity
requires a number of undesirable restrictions, which are discussed in
Section~\ref{related-work:modular-type-classes}.

\subsection{Modular implicits}
\label{intro:modular-implicits}

Taking inspiration from modular type classes and implicits, we propose a
system for ad-hoc polymorphism in OCaml based on passing implicit
\emph{module} parameters to functions based on their \emph{module
  type}. By basing our system on implicits, where a function's implicit
parameters must be given explicitly, we are able to avoid the
undesirable restrictions of modular type classes.  Fig.~\ref{fig:show}
demonstrates the \code{show} example written using our proposal.
\begin{figure}
\begin{lstlisting}[numbers=left]
    module type Show = sig
      type t
      val show : t -> string
    end

    let show {S : Show} x = S.show x

    implicit module Show_int = struct
      type t = int
      let show x = string_of_int x
    end

    implicit module Show_float = struct
      type t = float
      let show x = string_of_float x
    end

    implicit module Show_list {S : Show} = struct
      type t = S.t list
      let show x = string_of_list S.show x
    end

    let () =
      print_endline ("Show an int: " ^ show 5);
      print_endline ("Show a float: " ^ show 1.5);
      print_endline ("Show a list of ints: " ^ show [1; 2; 3]);
\end{lstlisting}
\caption{`Show` using modular implicits}
\label{fig:show}
\end{figure}

The \code{show} function (line 6) has two parameters: an implicit module
parameter \code{S} of module type \code{Show}, and an ordinary parameter
\code{x} of type \code{S.t}. When \code{show} is applied the module parameter
\code{S} does not need to be given explicitly. As with Scala implicits, when
this parameter is elided the system will search the modules which have been
made available for selection as implicit arguments for a module of the
appropriate type.

For example, on line 24, \code{show} is applied to \code{5}. This will
cause the system to search for a module of type \code{Show with type
  t = int}. Since \code{Show_int} is marked \code{implicit} and has the
desired type, it will be used as the implicit argument of \code{show}.

The \code{Show_list} module, defined on line 18, is an \emph{implicit
  functor} -- note the use of the \code{\{S : Show\}} syntax for its
parameter rather than the usual \code{(S : Show)} used for functor
arguments. This indicates that \code{Show_list} can be applied to create
implicit arguments, rather than used directly as an implicit argument.

For example, on line 26, \code{show} is applied to a list of
integers. This causes the system to search for an implicit module of
type \code{Show with type t = int list}. Such a module can be created by
applying the implicit functor \code{Show_list} to the implicit module
\code{Show_int}, so \code{Show_list(Show_int)} will be used as the implicit
argument.

Fig.~\ref{fig:monad} shows another example, illustrating how a simple library
for monads might look in our proposal.
\begin{figure}
\begin{lstlisting}[numbers=left]
    module type Monad = sig
      type +'a t
      val return : 'a -> 'a t
      val bind : 'a t -> ('a -> 'b t) -> 'b t
    end

    let return {M : Monad} x = M.return x

    let (>>=) {M : Monad} m k = M.bind m k

    let map {M : Monad} (m : 'a M.t) f =
      m >>= fun x -> return (f x)

    let join {M : Monad} (m : 'a M.t M.t) =
      m >>= fun x -> x

    let unless {M : Monad} p (m : unit M.t) =
      if p then return () else m

    implicit module Monad_option = struct
      type 'a t = 'a option
      let return x = Some x
      let bind m k =
        match m with
        | None -> None
        | Some x -> k x
    end

    implicit module Monad_list = struct
      type 'a t = 'a list
      let return x = [x]
      let bind m k = List.concat (List.map k m)
    end
\end{lstlisting}
\caption{`Monad` using modular implicits}
\label{fig:monad}
\end{figure}

The definitions of \code{map}, \code{join} and \code{unless} demonstrate
our proposal's support for higher-kinded polymorphism, analogous to
constructor classes in Haskell~\cite{DBLP:journals/jfp/Jones95}. This is
a more succinct form of higher-kinded polymorphism than is currently
available in OCaml's core language. Currently, higher-kinded
polymorphism is only supported directly using OCaml's verbose module
language or indirectly through an encoding based on
defunctionalisation~\cite{DBLP:conf/flops/YallopW14}.

The calls to \code{>>=} and \code{return} in the definitions of these
functions leave the module argument implicit. These cause the system to
search for a module of the appropriate type. In each case, the implicit module
parameter \code{M} of the function is selected because it has the
appropriate type and implicit module parameters are automatically made available
for selection as implicit arguments.

Like Scala's implicits, and unlike Haskell's type classes, our proposal
requires all of a function's implicit module parameters to be explicitly
declared. The \code{map} function (line 11) needs to be declared with
the module parameter \code{\{M : Monad\}} -- it could not be defined as follows:
\begin{lstlisting}
    let map m f =
      m >>= fun x -> return (f x)
\end{lstlisting}
because that would cause the system to try to resolve the implicit
module arguments to \code{>>=} and \code{return} to one of the implicit
modules available at the \emph{definition} of \code{map}. In this case,
this would result in an ambiguity error since either \code{Monad_option}
or \code{Monad_list} could be used.

\subsection{Contributions}
\label{intro:contributions}

The contributions of this paper are as follows.

\begin{itemize}
\item We introduce \textit{modular implicits}, a design for
  overloading centred around type-directed instantiation of implicit
  module arguments, that integrates harmoniously into a language with
  ML-style modules (Section~\ref{section:modular-implicits-design}).
  We show how to elaborate the extended language into standard OCaml,
  first by explicitly instantiating every implicit argument
  (Section~\ref{resolution}) and then by
  translating functions with implicit arguments into packages
  (Section~\ref{section:elaboration}).

\item The design of modular implicits involves only a small number of
  additions to the host language.  However, the close integration with
  the existing module language means that modular implicits naturally
  support a rich array of features, from constructs present in the
  original type classes proposal such as instance constraints
  (Section~\ref{section:instance-constraints}) and subclasses
  (Section~\ref{section:inheritance}) to extensions to the original
  type class proposal such as constructor classes
  (Section~\ref{section:constructor-classes}), multi-parameter
  type classes (Section~\ref{section:overloading-on-multiple-types}),
  associated types (Section~\ref{features:associated-types}) and
  backtracking (Section~\ref{resolution:backtracking}).  Further,
  modular implicits support a number of features not available with
  type classes.  For example, giving up canonicity -- without losing the
  motivating benefit of coherence (Section~\ref{section:canonicity}) --
  makes it possible to support local instances
  (Section~\ref{section:local-instances}), and basing resolution on
  module type inclusion results in a system in which a single instance
  can be used with a variety of different signatures
  (Section~\ref{section:structural-matching}).

\item Both resolution of implicit arguments and type inference involve a
  number of subtleties related to the interdependence of resolution and
  inference (Section~\ref{implicit-application:dependence}) and
  compositionality (Section~\ref{resolution:compositionality}).  We
  describe these at a high level here, leaving a more formal treatment
  to future work.

\item We have created a prototype implementation of our proposal based
  on OCaml 4.02.0. We describe some of the issues around implementing
  modular implicits (Section~\ref{implementation}).

\item Finally, we contextualise the modular implicits design within
  the wide body of related work, including Haskell type classes
  (Section~\ref{related-work:type-classes}), Scala implicits
  (Section~\ref{related-work:implicits}) canonical structures in Coq
  (Section~\ref{related-work:canonical-structures}), concepts in C++
  (Section~\ref{related-work:concepts}) and modular type classes in ML
  (Section~\ref{related-work:modular-type-classes}).
\end{itemize}

\section{The design of modular implicits}
\label{section:modular-implicits-design}

We present modular implicits as an extension to the OCaml language.
The OCaml module system includes a number of features, such as
first-class modules and functors, which make it straightforward to
elaborate modular implicits into standard OCaml.  However, the design
of modular implicits is not strongly tied to OCaml, and could be
integrated into similar languages in the ML family.

\subsection{New syntax}
\label{section:new-syntax}

Like several other designs for overloading based on implicit arguments,
modular implicits are based on three new features.
The first feature is a way to \textit{call} overloaded functions.
For example, we might wish to call an overloaded function \code{show},
implicitly passing a suitable value as argument, to convert an integer or a
list of floating-point values to a string.
The second feature is a way to \textit{abstract} overloaded functions.
For example, we might define a function \code{print} which calls \code{show}
to turn a value into a string in order to send it to standard output, but
which defers the choice of the implicit argument to pass to \code{show} to the
caller of \code{print}.
The third feature is a way to \textit{define} values that can be used as
implicit arguments to overloaded functions.
For example, we might define a family of modules for building string
representations for values of many different types, suitable for passing as
implicit arguments to \code{show}.

Figure~\ref{figure:syntax} shows the new syntactic forms for modular
implicits, which extend the syntax of OCaml 4.02~\cite{leroy:hal-00930213}.

There is one new form for types,
\begin{lstlisting}[style=ocamlgrammar]
    { M : T } -> t
\end{lstlisting}
which makes it possible to declare \code{show} as a function with an
implicit parameter \code{S} of module type \code{Show}, a second
parameter of type \code{S.t}, and the return type \code{string}:
\begin{lstlisting}
    val show : {S: Show} -> S.t -> string
\end{lstlisting}
or to define \code{+} as a function with an implicit parameter \code{N}
of module type \code{Num}, two further parameters of type \code{N.t},
and the return type \code{N.t}:
\begin{lstlisting}
    val ( + ) : {N: Num} -> N.t -> N.t -> N.t
\end{lstlisting}
There is a new kind of parameter for constructing functions with
implicit arguments:
\begin{lstlisting}[style=ocamlgrammar]
    { M : T }
\end{lstlisting}
The following definition of \code{show} illustrates the use of
implicit parameters:
\begin{lstlisting}
    let show {S : Show} (v : S.t) = S.show v
\end{lstlisting}
The braces around the \code{S : Show} indicate that \code{S} is an implicit
module parameter of type \code{Show}.  The type \code{Show} of \code{S} is a
standard OCaml module type, which might be defined as in
Figure~\ref{fig:show}.

There is also a new kind of argument for calling functions with
implicit arguments:
\begin{lstlisting}[style=ocamlgrammar]
    { M }
\end{lstlisting}
For example, the \code{show} function might be called as follows using this
argument syntax:
\begin{lstlisting}
    show {Show_int} 3
\end{lstlisting}
This is an explicitly-instantiated implicit application.  Calls to \code{show}
can also omit the first argument, leaving it to be supplied by a
resolution procedure (described in Section~\ref{resolution}):
\begin{lstlisting}
    show 3
\end{lstlisting}
Implicit application requires that the function have non-module
parameters after the module parameter -- implicit application is
indicated by providing arguments for these later parameters without
providing a module argument for the module parameter. This approach
simplifies type-inference and is in keeping with how OCaml handles
optional arguments. It also ensures that all function applications,
which may potentially perform side-effects, are syntactically function
applications.

There are two new declaration forms.  Here is the first, which
introduces an implicit module:
\begin{lstlisting}[style=ocamlgrammar]
    implicit module M LSTAR{M$_i$ : T$_i$}RSTAR = E
\end{lstlisting}
Implicit modules serve as the implicit arguments to overloaded
functions like \code{show} and \code{+}.  For example, here is the
definition of an implicit module \code{Show_int} with two members: a
type alias \code{t} and a value member \code{show} which uses the
standard OCaml function \code{string_of_int}
\begin{lstlisting}
    implicit module Show_int = struct
      type t = int
      let show = string_of_int
    end
\end{lstlisting}
Implicit modules can themselves have implicit parameters.  For example, here
is the definition of an implicit module \code{Show_list} with an implicit
parameter which also satisfies the \code{Show} signature:
\begin{lstlisting}
    implicit module Show_list {A: Show} = struct
      type t = A.t list
      let show l =
        "["^ String.concat ", " (List.map A.show l) ^"]"
    end
\end{lstlisting}
Implicit modules with implicit parameters are called \emph{implicit functors}.
Section~\ref{resolution} outlines how implicit modules are selected for use as
implicit arguments.

The second new declaration form brings implicit modules into scope, making
them available for use in resolution:
\begin{lstlisting}[style=ocamlgrammar]
    open implicit M
\end{lstlisting}
For example, the declaration
\begin{lstlisting}
    open implicit List
\end{lstlisting}
makes every implicit module bound in the module \code{List} available to the
resolution procedure in the current scope.

There are also local versions of both declaration forms, which bind a module
or bring implicits into scope within a single expression:
\begin{lstlisting}[style=ocamlgrammar]
    let implicit module M LSTAR{M$_i$ : T$_i$}RSTAR = E in e
    let open implicit M in e
\end{lstlisting}

Implicit module declarations, like other OCaml declarations, bind names within
modules, and so the signature language must be extended to support implicit
module descriptions.  There are two new forms for describing implicit modules
in a signature:
\begin{lstlisting}[style=ocamlgrammar]
    implicit module M LSTAR{M$_i$ : T$_i$}RSTAR : T
    implicit module M LSTAR{M$_i$ : T$_i$}RSTAR = M
\end{lstlisting}
The first form describes a binding for an implicit module by means of its
type.  For example, here is a description for the module \code{Show_list}:
\begin{lstlisting}
    implicit module Show_int : Show with type t = int
\end{lstlisting}
The second form describes a binding for an implicit module by means of an
equation~\cite{module-aliases}.  For example, here is a description for a
module \code{S}, which is equal to \code{Show_int}
\begin{lstlisting}
    implicit module S = Show_int
\end{lstlisting}

\begin{figure}
\textbf{Types}


\begin{lstlisting}[style=ocamlgrammar]
typexpr DEF $\ldots$ | { module-name : module-type } -> typexpr
\end{lstlisting}

\textbf{Expressions}


\begin{lstlisting}[style=ocamlgrammar]
parameter DEF $\ldots$ $\mid$ { module-name : module-type }
\end{lstlisting}

\begin{lstlisting}[style=ocamlgrammar]
argument DEF $\ldots$ $\mid$ { module-expr }
\end{lstlisting}


\begin{lstlisting}[style=ocamlgrammar]
expr DEF $\ldots$
$\mid$ let implicit module module-name LSTAR{module-name : module-type}RSTAR =
    module-expr in expr
$\mid$ let open implicit module-path in expr
\end{lstlisting}

\textbf{Bindings and declarations}


\begin{lstlisting}[style=ocamlgrammar]
definition DEF $\ldots$
$\mid$ implicit module module-name LSTAR{module-name : module-type}RSTAR = module-expr
$\mid$ open implicit module-path
\end{lstlisting}

\textbf{Signature declarations}


\begin{lstlisting}[style=ocamlgrammar]
specification DEF $\ldots$
$\mid$ implicit module module-name LSTAR{module-name : module-type}RSTAR : module-type
$\mid$ implicit module module-name LSTAR{module-name : module-type}RSTAR = module-path
\end{lstlisting}

\caption{\label{figure:syntax}Syntax for modular implicits}
\end{figure}

\subsection{Resolving implicit arguments}
\label{resolution}

As we saw in Section~\ref{section:new-syntax}, a function which accepts an
implicit argument may receive that argument either implicitly or explicitly.
The \textit{resolution} process removes implicit arguments by replacing them
with explicit arguments constructed from the modules in the implicit search
space.

Resolving an implicit argument \code{M} involves two steps.  The first
step involves gathering constraints -- that is, equations on
types\footnote{Constraints on module types and module aliases are also
  possible, but we leave them out of our treatment} within \code{M}
-- based on the context in which the application appears.  For
example, the application
\begin{lstlisting}
    show 5
\end{lstlisting}
should generate a constraint
\begin{lstlisting}
    S.t = int
\end{lstlisting}
on the implicit module argument \code{S} passed to \code{show}.  The
second step involves searching for a module which satisfies the
constrained argument type.  Resolving the implicit argument for the
application \code{show 5} involves searching for a module \code{S} with
the type
\begin{lstlisting}
    Show
\end{lstlisting}
that also satisfies the constraint
\begin{lstlisting}
    S.t = int
\end{lstlisting}
The following sections consider these steps in more detail.

\subsubsection{Generating argument constraints}
\label{implicit-application:generating}

Generating implicit argument constraints for an application \code{f x}
with an implicit argument \code{M} of type \code{S} involves building
a substitution which equates each type \code{t} in \code{S} with a
fresh type variable \code{'a}, then using unification to further
constrain \code{'a}.  For example, \code{show} has type:
\begin{lstlisting}
    {S : Show} -> S.t -> string
\end{lstlisting}
and the module type \code{Show} contains a single type \code{t}.
The constraint generation procedure generates the constraint
\begin{lstlisting}
   S.t = 'a
\end{lstlisting}
for the implicit parameter, and refines the remainder of the type of
\code{show} to
\begin{lstlisting}
    'a -> string
\end{lstlisting}
Type-checking the application \code{show 5} using this type reveals
that \code{'a} should be unified with \code{int}, resulting in the
following constraint for the implicit parameter:
\begin{lstlisting}
    S.t = int
\end{lstlisting}

In our treatment we assume that all implicit arguments have structure
types.  However, functor types can also be supported by introducing
similar constraints on the results of functor applications.

Generating implicit argument constraints for higher-kinded types
involves some additional subtleties compared to generating constraints
for basic types.  With higher-kinded types, type constructors cannot
be directly replaced by a type variable, since OCaml does not support
higher-kinded type variables.  Instead, each application of a
parameterised type constructor must be replaced by a separate type
variable.

For example, the \code{map} function has the following type:
\begin{lstlisting}
    {M : Monad} -> 'a M.t -> ('a -> 'b) -> 'b M.t
\end{lstlisting}
After substituting out the module parameter, the type becomes:
\begin{lstlisting}
    'c -> ('a -> 'b) -> 'd
\end{lstlisting}
with the following constraints:
\begin{lstlisting}
    'a M.t = 'c
    'b M.t = 'd
\end{lstlisting}
Type-checking a call to \code{map} determines the type variables
\code{'c} and \code{'d}.  For example, the following call to \code{map}:
\begin{lstlisting}
    let f x =
      map [x; x] (fun y -> (y, y))
\end{lstlisting}
refines the constraints to the following:
\begin{lstlisting}
    'a M.t = 'e list
    ('a * 'a) M.t = 'd
\end{lstlisting}
where \code{'a}, \code{'d} and \code{'e} are all type variables
representing unknown types.

We might be tempted to attempt to refine the constraints further, inferring
that \code{'a = 'e} and that \code{s M.t = s list} for any type
\code{s}. However, this inference is not necessarily correct. If, instead of
\code{Monad_list}, the following module was in scope:
\begin{lstlisting}
    implicit module Monad_odd = struct
      type 'a t = int list
      let return x = [1; 2; 3]
      let bind m f = [4; 5; 6]
    end
\end{lstlisting}
then those inferences would be incorrect. Since the definition of the
type \code{t} in \code{Monad_odd} simply discards its parameter, there is
no requirement for \code{'e} to be equal to \code{'a}.  Further, for any
type \code{s}, \code{s Monad_odd.t} would be equal to \code{int list},
not to \code{s list}.

In fact, inferring additional information from these constraints before
performing resolution would constitute second-order unification, which
is undecidable in general. Resolution does not require second-order
unification as it only searches amongst possible solutions rather than
finding a most general solution.

Once the constraints have been used to resolve the module argument
\code{M} to \code{Monad_list}, we can safely substitute \code{list} for
\code{M.t} which gives us the expected type equalities.

\subsubsection{Searching for a matching module}
\label{resolution:modules}

Once the module type of the implicit argument has been constrained, the
next step is to find a suitable module.  A module is considered suitable
for use as the implicit argument if it satisfies three criteria:
\begin{enumerate}
\item It is constructed from the modules and functors in the implicit search
space.
\item It matches the constrained module type for the implicit
argument.
\item It is unique -- that is, it is the only module satisfying the first two criteria.
\end{enumerate}

\paragraph{The implicit search space}

The implicit search space consists of those modules which have been bound with
\code{implicit module} or \code{let implicit module}, or which are in scope as
implicit parameters.  For example, in the following code all of \code{M},
\code{P} and \code{L} are included in the implicit search space at the point
of the expression \code{show v}
\begin{lstlisting}
    implicit module M = M1
    module N = M2
    let f {P : Show} v ->
      let implicit module L = M3 in show v
\end{lstlisting}


Furthermore, in order to avoid unnecessary ambiguity, resolution is
restricted to those modules which are accessible using unqualified
names.  An implicit sub-module \code{M} in a module \code{N} is not in
scope unless \code{N} has been opened.  Implicit modules from other
modules can be brought into scope using \code{open implicit} or
\code{let open implicit}.

\paragraph{Module type matching}

Checking that an implicit module \code{M} matches an implicit argument type
involves checking that the signature of \code{M} matches the signature of the
argument and that the constraints generated by type checking hold for
\code{M}.
As with regular OCaml modules, signature matching allows \code{M} to have more
members than the argument signature.  For example, the following module
matches the module type \code{Show with type t = int}, despite the fact that
the module has an extra value member, \code{read}:
\begin{lstlisting}
    implicit module Show_read_int = struct
      type t = int
      let show = string_of_int
      let read = int_of_string
    end
\end{lstlisting}
%
Constraint matching is defined in terms of substitution: can the type
variables in the generated constraint set be instantiated such that the
equations in the set hold for \code{M}?  For example, \code{Monad_list} meets
the constraint
\begin{lstlisting}
    'a M.t = int list
\end{lstlisting}
by replacing \code{'a} with \code{int}, giving the valid equation
\begin{lstlisting}
    int Monad_list.t = int list
\end{lstlisting}

In simple cases, resolution is simply a matter of trying each implicit module
in turn to see whether it matches the signature and generated constraints.

However, when there are implicit functors in scope the resolution procedure
becomes more involved. For example, the declaration for \code{Show_list} from
Figure~\ref{fig:show} allows modules such as \\
\code{Show_list(Show_int)} to be used as implicit module arguments:
\begin{lstlisting}
    implicit module Show_list {S : Show} = struct
      type t = S.t list
      let show l = string_of_list S.show l
    end
\end{lstlisting}

Checking whether an implicit functor can be used to create a module
which satisfies an implicit argument's constraints involves substituting
an application of the functor for the implicit argument and checking
that the equations hold. For example, applying \code{Show_list} to create
a module \code{M} could meet the constraint:
\begin{lstlisting}
    M.t = int list
\end{lstlisting}
as substituting an application of the functor gives:
\begin{lstlisting}
    Show_list(S).t = int list
\end{lstlisting}
which expands out to
\begin{lstlisting}
    S.t list = int list
\end{lstlisting}
This generates a constraint on the argument \code{S} to \code{Show_list}:
\begin{lstlisting}
    S.t = int
\end{lstlisting}
Since \code{Show_int} satisfies this new constraint,
\code{Show_list(Show_int)} meets the original constraint.

Matching of candidate implicit arguments against signatures is defined
in terms of OCaml's signature matching relation, and so it supports
the full OCaml signature language, including module types, functors,
type definitions, exception specifications, and so on.  However, since
modular implicits extend the signature language with new constructs
(Figure~\ref{figure:syntax}), the matching relation must also be
extended.  Matching for function types with implicit parameters is
straightforward, and corresponds closely to matching for function
types with labeled arguments.  In particular, matching for function
types does not permit elision or reordering of implicit parameters.

\paragraph{Uniqueness}

In order to maintain coherence, modular implicits require the module returned
by resolution to be unique.  Without a uniqueness requirement the result of
resolution (and hence the behaviour of the program) might depend on some
incidental aspect of type-checking.

To check uniqueness all possible solutions to a resolution must be
considered. This requires that the search for possible resolutions
terminate: if the resolution procedure does not terminate then we do not
know whether there may be multiple solutions.

The possibility of non-termination and the interdependence between
resolution and type inference
(Section~\ref{implicit-application:dependence}) mean that checking
uniqueness of solutions is incomplete, and can report ambiguity
errors in cases which are not actually ambiguous. As with similar forms
of incomplete inference, our proposal aims to make such cases
predictable by using simple specifications of the termination conditions
and of the permitted dependencies between resolution and type inference.

\paragraph{Termination}

Implicit functors can be used multiple times whilst resolving a single
implicit argument.  For example
\begin{lstlisting}
    show [ [1; 2; 3]; [4; 5; 6] ]
\end{lstlisting}
will resolve the implicit argument of \code{show} to
\code{Show_list(Show_list(Show_int))}.

This means that care is needed to avoid non-termination in the resolution
procedure. For example, the following functor, which tries to define how to
show a type in terms of how to show that type, is obviously not well-founded:
\begin{lstlisting}
    implicit module Show_it {S : Show} = struct
      type t = S.t
      let show = S.show
    end
\end{lstlisting}

Type classes ensure the termination of resolution through a number of
restrictions on instance declarations. However, termination of an implicit
parameter resolution depends on the scope in which the resolution is
performed. For this reason, the modular implicits system places restrictions
on the behaviour of the resolution directly and reports an error only when a
resolution which breaks these restrictions is actually attempted.

When considering a module expression containing multiple applications
of an implicit functor, such as the following:
\begin{lstlisting}
    Show_list(Show_list( ... ))
\end{lstlisting}
the system checks that the constraints that each application of the functor
must meet are strictly smaller than the previous application of the
functor. ``Strictly smaller'' is defined point-wise: all constraints must be
smaller, and at least one constraint must be strictly smaller.

For example, resolving an implicit module argument \code{S} of type
\code{Show} with the following constraint
\begin{lstlisting}
    S.t = int list list
\end{lstlisting}
would involve considering \code{Show_list(Show_list(T))} where \code{T} is
not yet determined. The first application of \code{Show_list} would
generate a constraint on its argument \code{R}:
\begin{lstlisting}
    R.t = int list
\end{lstlisting}
Thus the second application of \code{Show_list} must meet constraints
which are strictly smaller than the constraints which the first
application of \code{Show_list} met, and resolution can safely continue.

Whereas, considering \code{Show_it(Show_it(S))} for the same
constraint, the first application of \\
\code{Show_it} would generate a constraint on its argument \code{R}:
\begin{lstlisting}
    R.t = int list list
\end{lstlisting}
Thus the second application of \code{Show_it} must meet constraints which
are the same as the constraints which the first application of
\code{Show_it} met, and resolution would fail with a termination error.

Multiple applications of a functor are not necessarily successive,
since there may be other applications between them. For example, the
expression to be checked could be of the form:
\begin{lstlisting}
    Show_this(Show_that(Show_this( ... )))
\end{lstlisting}
In this case, the ``strictly smaller'' condition applies between the
outer application of \code{Show_this} and the inner application of
\code{Show_this}. The application of \code{Show_that} will not be compared
to the applications of \code{Show_this}.

As termination is required to check uniqueness, failure to meet the
termination restrictions must be treated as an error.  The system cannot simply
ignore the non-terminating possibilities and continue to look for an
alternative resolution.

\subsection{Elaboration}
\label{section:elaboration}


Once all implicit arguments in a program have been instantiated there
is a phrase-by-phrase elaboration which turns each new construct into a
straightforward use of existing OCaml constructs.  The elaboration
makes use of OCaml's first-class modules (packages), turning functions
with implicit arguments into first-class functors,

\begin{figure}

\paragraph{Types}
The type
\begin{lstlisting}
    {M: S} -> t
\end{lstlisting}
elaborates to the package type
\begin{lstlisting}
    (module functor (M:S) -> sig val value : t end)
\end{lstlisting}

\paragraph{Abstractions}

The abstraction expression
\begin{lstlisting}
    fun {M: S} -> e
\end{lstlisting}
of type
\begin{lstlisting}
    {M: S} -> t
\end{lstlisting}
elaborates to the package expression
\begin{lstlisting}
    (module functor (M: S) -> struct
       let value = e
     end)
\end{lstlisting}
of type
\begin{lstlisting}
   (module functor (M: S) -> sig val value : t end))
\end{lstlisting}

\paragraph{Applications}

The application expression
\begin{lstlisting}
    f {M}
\end{lstlisting}
elaborates to the expression
\begin{lstlisting}
    let module F = (val f) in
    let module R = F(M) in
      R.value
\end{lstlisting}

\paragraph{Bindings and declarations}

The implicit module binding
\begin{lstlisting}[mathescape=true]
    implicit module M { M1 : T1 } { M2 : T2 } $\ldots$ { Mn : Tn } = N
\end{lstlisting}
elaborates to the expression
\begin{lstlisting}[mathescape=true]
    module M (M1 : T1) (M2 : T2) $\ldots$ (Mn : Tn) = N
\end{lstlisting}
(and similarly for local bindings and signatures).

The statement
\begin{lstlisting}[mathescape=true]
    open implicit M
\end{lstlisting}
is removed from the program (and similarly for local \code{open implicit}
bindings).

\caption{\label{figure:elaboration}Elaboration from a
  fully-instantiated program into OCaml}
\end{figure}

Figure~\ref{figure:elaboration} gives the elaboration from a
fully-instantiated program into implicit-free OCaml.
The types of functions which accept implicit arguments
\begin{lstlisting}
   {M: S} -> t
\end{lstlisting}
become first-class functor types
\begin{lstlisting}
   (module functor (M:S) -> sig val value : t end)
\end{lstlisting}
with a functor parameter in place of the implicit parameter \code{M}
and a signature with a single value member of type \code{t} in place
of the return type \code{t}.  (The syntax used here for the
first-class functor type is not currently accepted by OCaml, which
restricts the types of first-class modules to named module types, but
the restriction is for historical reasons only, and so we ignore it in
our treatment.  The other parts of the elaboration target entirely
standard OCaml.)

An expression which constructs a function that accepts an implicit argument
\begin{lstlisting}
   fun {M: S} -> e
\end{lstlisting}
becomes an expression which packs a functor
\begin{lstlisting}
   (module functor (M: S) -> struct
      let value = e
    end)
\end{lstlisting}
following the elaboration on types, turning the implicit argument into
a functor argument and the body into a single value binding
\code{value}.

The applications of a function \code{f} to an instantiated implicit
arguments \code{M}
\begin{lstlisting}
    f {M}
\end{lstlisting}
becomes an expression which unpacks \code{f} as a functor \code{F},
applies \code{F} to the module argument \code{M}, and projects the
value component from the result:
\begin{lstlisting}
    let module F = (val f) in
    let module R = F(M) in
      R.value
\end{lstlisting}
Care must, of course, be taken to ensure that the name \code{F} does not
collide with any of the free variables in the module expression
\code{M}.

Each implicit module binding
\begin{lstlisting}[mathescape=true]
   implicit module M { M1 : T1 } { M2 : T2 } $\ldots$ { Mn : Tn } = N
\end{lstlisting}
becomes under the elaboration a binding for a regular module, turning
implicit parameters into functor parameters:
\begin{lstlisting}[mathescape=true]
   module M (M1 : T1) (M2 : T2) $\ldots$ (Mn : Tn) = N
\end{lstlisting}
The implicit module binding for \code{M} introduces \code{M} both into
the implicit search space and the standard namespace of the program.
The implicit search space is not used in the program after
elaboration, and so the elaborated binding introduces \code{M} only
into the standard namespace.  The elaboration for local bindings and
signatures is the same, mutatis mutandis.

The statement
\begin{lstlisting}[mathescape=true]
    open implicit M
\end{lstlisting}
serves no purpose after elaboration, and so the elaboration simply
removes it from the program.  Similarly, the statement
\begin{lstlisting}[mathescape=true]
    let open implicit M in e
\end{lstlisting}
is elaborated simply to the body:
\begin{lstlisting}
    e
\end{lstlisting}

\subsection{Why target first-class functors?}
\label{module-parameters}

The elaboration from an instantiated program into first-class functors
is quite simple, but the syntax of implicit arguments suggests an even
simpler translation which turns each function with an
implicit parameter into a function (rather than a functor) with
a first-class module parameter.  For example, here is the
definition of \code{show} once again:
\begin{lstlisting}
   let show {S : Show} (v : S.t) = S.show v
\end{lstlisting}
Under the elaboration in Figure~\ref{figure:elaboration} the
definition of \code{show} becomes the following first-class functor
binding:
\begin{lstlisting}
   let show =
     (functor (S: Show) -> struct
         let value = fun (v : S.t) -> S.show v
       end)
\end{lstlisting}
but we could instead elaborate into a function with a first-class module
argument
\begin{lstlisting}
   let show (module S: Show) (v : S.t) = S.show v
\end{lstlisting}
of type
\begin{lstlisting}
   (module Show with type t = 'a) -> 'a -> string
\end{lstlisting}
Similarly, under the elaboration in Figure~\ref{figure:elaboration}
the application of \code{show} to an argument
\begin{lstlisting}
   show {Show_int}
\end{lstlisting}
is translated to an expression with two local module bindings, a
functor application and a projection:
\begin{lstlisting}
   let module F = (val show) in
   let module R = F(Show_int) in
     R.value
\end{lstlisting}
but under the elaboration into functions with first-class module
arguments the result is a simple application of \code{show} to a packed module:
\begin{lstlisting}
   show (module Show_int)
\end{lstlisting}

However, the extra complexity in targeting functors rather than
functions pays off in support for higher-rank and higher-kinded
polymorphism.

\subsubsection{Higher-rank polymorphism}
It is convenient to have overloaded functions be first-class citizens
in the language.  For example, here is a function which takes an
overloaded function \code{sh} and applies it both to an integer and to
a string:
\begin{lstlisting}
    let show_stuff (sh : {S : Show} -> S.t -> string) =
      (sh {Show_int} 3, sh {Show_string} "hello")
\end{lstlisting}
This application of the parameter \code{sh} at two different types
requires \code{sh} to be polymorphic in the type \code{S.t}.  This
form of polymorphism, where function arguments themselves can be
polymorphic functions, is sometimes called \textit{higher-rank}
polymorphism.

The elaboration of overloaded functions into first-class functors
naturally supports higher-rank polymorphism, since functors themselves
can behave like polymorphic functions, with type members in their
arguments.  Here is the elaboration of \code{show_stuff}:
\begin{lstlisting}
    let show_stuff (sh : (module functor (S : Show) -> sig
                            val value : S.t -> string
                          end)) =
      let module F1 = (val sh) in
      let module R1 = F1(Show_int) in
      let module F2 = (val sh) in
      let module R2 = F2(Show_string) in
        (R1.value 5, R2.value "hello")
\end{lstlisting}
The two functor applications \code{F1(Show_int)} and
\code{F2(Show_string)} correspond to two instantiations of a
polymorphic function.

In contrast, if we were to elaborate overloaded functions into
ordinary functions with first-class module parameters then the result
of the elaboration would not be valid OCaml.  Here is the result of
such an elaboration:
\begin{lstlisting}
    let show_stuff (sh : (module S with type t = 'a)
                            -> 'a -> string) =
      sh (module Show_int) 3 ^ " "
        ^ sh (module Show_string) "hello"
\end{lstlisting}
Since \code{sh} is a regular function parameter, OCaml's type rules
assign it a monomorphic type.  The function is then rejected, because
\code{sh} is applied to modules of different types within the body.

\subsubsection{Higher-kinded polymorphism}

First-class functors also provide support for \textit{higher-kinded}
polymorphism -- that is, polymorphism in type constructors which have
parameters.  For example, Figure~\ref{fig:monad} defines a number of
functions that are polymorphic in the monad on which they operate,
such as \code{map}, which has the following type:
\begin{lstlisting}
    val map : {M : Monad} -> 'a M.t -> ('a -> 'b) -> 'b M.t
\end{lstlisting}
This type is polymorphic in the parameterised type constructor
\code{M.t}.

Once again, elaborating overloaded functions into first-class functors
naturally supports higher-kinded polymorphism, since functor arguments
can be used to abstract over parameterised type constructors.  Here is
the definition of \code{map} once again:
\begin{lstlisting}
    let map {M : Monad} (m : 'a M.t) f =
      m >>= fun x -> return (f x)
\end{lstlisting}
and here its its elaboration:
\begin{lstlisting}
    let map =
      (functor (M: Monad) -> struct
         let value =
           let module F_bind = (val (>>=)) in
           let module R_bind = F_bind(M) in
           let module F_ret = (val return) in
           let module R_ret = F_ret(M) in
             R_bind.value m (fun x -> R_ret (f x))
       end)
\end{lstlisting}
As with higher-rank polymorphism, there is no suitable elaboration of
overloaded functions involving higher-kinded polymorphism into
functions with first-class module parameters, since higher-kinded
polymorphism is not supported in OCaml's core language.

\subsubsection{First-class functors and type inference}

Type inference for higher-rank and full higher-kinded polymorphism is
undecidable in the general case, and so type systems which support such
polymorphism require type annotations.  For instance, annotations are
required on all first-class functor parameters, and on recursive
definitions of recursive functors. The same requirements apply to
functions with implicit module arguments.

For example, the following function will not type-check if the \code{sh}
parameter is not annotated with its type:
\begin{lstlisting}
    let show_three sh =
      sh {Show_int} 3
\end{lstlisting}
Instead, \code{show_three} must be defined as follows:
\begin{lstlisting}
    let show_three (sh : {S : Show} -> S.t -> string)  =
      sh {Show_int} 3
\end{lstlisting}

Requiring type annotations means that type inference is not \emph{order
  independent} -- if the body of \code{show\_three} were type-checked
before its parameter list then inference would fail. To maintain
predictability of type inference, some declarative guarantees are made
about the order of type-checking; for example, a variable's binding will
always be checked before its uses. If type inference of a program only
succeeds due to an ordering between operations which is not ensured by
these guarantees then the OCaml compiler will issue a warning.

\section{Modular implicits by example}
\label{section:modular-implicits-by-example}

The combination of the implicit resolution mechanism and the integration
with the module language leads to a system which can support a wide
range of programming patterns. We demonstrate this with a selection of
example programs.

\subsection{Defining overloaded functions}
\label{section:overloaded-functions}

Some overloaded functions, such as \code{show} from Figure~\ref{fig:show},
simply project a member of the implicit module argument.  However, it is also
common to define an overloaded function in terms of an existing overloaded
function.  For example, the following \code{print} function composes the
standard OCaml function \code{print_string} with the overloaded function
\code{show} to print a value to standard output:
\begin{lstlisting}
    let print {X: Show} (v: X.t) =
      print_string (show v)
\end{lstlisting}
It is instructive to consider the details of resolution for the call to
\code{show} in the body of \code{print}.  As described in
Section~\ref{resolution}, resolution of the implicit argument \code{S}
of \code{show} involves generating constraints for the types in
\code{S}, unifying with the context to refine the constraints, and then
searching for a module \code{M} which matches the signature of
\code{Show} and satisfies the constraints.

Since there is a single type \code{t} in the signature \code{Show}, resolution
begins with the constraint set
\begin{lstlisting}
   S.t = 'a
\end{lstlisting}
and gives the variable \code{show} the type \code{'a -> string}.  Unification
with the ascribed type of the parameter \code{v} instantiates \code{'a},
refining the constraint
\begin{lstlisting}
   S.t = X.t
\end{lstlisting}
Since the type \code{X.t} is an abstract member of the implicit module
parameter \code{X}, the search for a matching module returns \code{X} as the
unique implicit module which satisfies the constraint.

The ascription on the parameter \code{v} plays an essential role in this
process.  Without the ascription, resolution would involve searching for an
implicit module of type \code{Show} satisfying the constraint \code{S.t = 'a}.
Since any implicit module matching the signature \code{Show} satisfies this
constraint, regardless of the definition of \code{t}, the resolution procedure
will fail with an ambiguity error if there are multiple implicit modules in
scope matching \code{Show}.

\subsection{Instance constraints}
\label{section:instance-constraints}

Haskell's instance constraints make it possible to restrict the set of
instantiations of type parameters when defining overloaded functions.  For
example, here is an instance of the \code{Show} class for the pair constructor
\code{(,)}, which is only available when there are also an instance of
\code{Show} for the type parameters \code{a} and \code{b}:
\begin{lstlisting}[style=haskell]
    instance (Show a, Show b) => Show (a, b) where
       show (x, y) = "(" ++ show x ++ "," ++ show y ++ ")"
\end{lstlisting}
With modular implicits, instance constraints become parameters to implicit
functor bindings:
\begin{lstlisting}
    implicit module Show_pair {A: Show} {B: Show} = struct
       type t = A.t * B.t
       let show (x, y) = "(" ^ A.show x ^ "," ^ B.show y ^ ")"
    end
\end{lstlisting}
It is common for the types of implicit functor parameters to be related to the
type of the whole, as in this example, where the parameters each match
\code{Show} and the result has type \code{Show with type t =} \code{A.t * B.t}.
However, neither instance constraints nor implicit module parameters require
that the parameter and the result types are related.  Here is the definition
of an implicit module \code{Complex_cartesian}, which requires only that the parameters
have implicit module bindings of type \code{Num}, not of type \code{Complex}:
\begin{lstlisting}
    implicit module Complex_cartesian {N: Num} = struct
      type t = N.t complex_cartesian
      let conj { re; im } = { re; im = N.negate im }
    end
\end{lstlisting}
(We leave the reader to deduce the definitions of the \code{complex_cartesian}
type and of the \code{Num} signature.)

\subsection{Inheritance}
\label{section:inheritance}

Type classes in Haskell provide support for \emph{inheritance}. For
example, the \code{Ord} type class is defined as inheriting from the
\code{Eq} type class:
\begin{lstlisting}[style=haskell]
    class Eq a where
      (==) :: a -> a -> Bool

    class Eq a => Ord a where
      compare :: a -> a -> Ordering
\end{lstlisting}

This means that instances of \code{Ord} can only be created for types
which have an instance of \code{Eq}. By declaring \code{Ord} as
inheriting from \code{Eq}, functions can use both \code{==} and
\code{compare} on a type with a single constraint that the type have an
\code{Ord} instance.

\subsubsection{The ``diamond'' problem}

It is tempting to try to implement inheritance with modular implicits by
using the structural subtyping provided by OCaml's modules. For example,
one might try to define \code{Ord} and \code{Eq} as follows:
\begin{lstlisting}
    module type Eq = sig
      type t
      val equal : t -> t -> bool
    end

    let equal {E : Eq} x y = E.equal x y

    module type Ord = sig
      type t
      val equal : t -> t -> bool
      val compare : t -> t -> int
    end

    let compare {O : Ord} x y = O.compare x y
\end{lstlisting}
which ensures that any module which can be used as an implicit
\code{Ord} argument can also be used as an implicit \code{Eq}
argument. For example, a single module can be created for both equality
and comparison of integers:
\begin{lstlisting}
    implicit module Ord_int = struct
      type t = int
      let equal = Int.equal
      let compare = Int.compare
    end
\end{lstlisting}

However, an issue arises when trying to implement implicit functors for
type constructors using this scheme. For example, we might want to
define the following two implicit functors:
\begin{lstlisting}
    implicit module Eq_list {E : Eq} = struct
      type t = E.t list
      let equal x y = List.equal E.equal x y
    end

    implicit module Ord_list {O : Ord} = struct
      type t = O.t list
      let equal x y = List.equal O.equal x y
      let compare x y = List.compare O.compare x y
    end
\end{lstlisting}
which implement \code{Eq} for lists of types which implement \code{Eq},
and implement \code{Ord} for lists of types which implement \code{Ord}.

The issue arises when we wish to resolve an \code{Eq} instance for a
list of a type which implements \code{Ord}. For example, we might wish
to apply the \code{equal} function to lists of ints:
\begin{lstlisting}
    equal [1; 2; 3] [4; 5; 6]
\end{lstlisting}
The implicit argument in this call is ambiguous: we can use either
\code{Eq_list(Ord_int)} or \\ \code{Ord_list(Ord_int)}.

This is a kind of ``diamond'' problem: we can restrict \code{Ord_int} to an
\code{Eq} and then lift it using \code{Eq_list}, or we can lift
\code{Ord_int} using \code{Ord_list} and then restrict the result to an
\code{Eq}.

In Haskell, the problem is avoided by canonicity -- it doesn't matter
which way around the diamond we go, we know that the result will be the
same.

\subsubsection{Module aliases}

OCaml provides special support for \emph{module
  aliases}~\cite{module-aliases}. A module can be defined as an alias for
another module:
\begin{lstlisting}
    module L = List
\end{lstlisting}
This defines a new module whose type is the singleton type ``\code{=
  List}''. In other words, the type of \code{L} guarantees that it is
equal to \code{List}. This equality allows types such as
\code{Set(List).t} and \code{Set(L).t} to be considered equal.

Since \code{L} is statically known to be equal to \code{List}, we do not
consider an implicit argument to be ambiguous if \code{L} and
\code{List} are the only possible choices.

In our proposal we extend module aliases to support implicit functors.
For example,
\begin{lstlisting}
    implicit module Show_l {S : Show} = Show_list{S}
\end{lstlisting}
creates a module alias. This means that \code{Show_l(Show_int)} is an
alias for \code{Show_list(Show_int)}, and its type guarantees that the two
modules are equal.

In order to maintain coherence we must require that all implicit
functors be pure. If \code{Show_list} performed side-effects then two
separate applications of it would not necessarily be equal. We ensure
this using the standard OCaml value restriction. This is a very
conservative approximation of purity, but we do not expect it to be too
restrictive in practice.

\subsubsection{Inheritance with module aliases}

Using module aliases we can implement inheritance using modular
implicits. Our \code{Ord} example is encoded as follow:
\begin{lstlisting}
    module type Eq = sig
      type t
      val equal : t -> t -> bool
    end

    let equal {E : Eq} x y = E.equal x y

    module type Ord = sig
      type t
      module Eq : Eq with type t = t
      val compare : t -> t -> int
    end

    let compare {O : Ord} x y = O.compare x y

    implicit module Eq_ord {O : Ord} = O.Eq

    implicit module Eq_int = struct
      type t = int
      let equal = Int.equal
    end

    implicit module Ord_int = struct
      type t = int
      module Eq = Eq_int
      let compare = Int.compare
    end

    implicit module Eq_list {E : Eq} = struct
      type t = E.t list
      let equal x y = List.equal E.equal x y
    end

    implicit module Ord_list {O : Ord} = struct
      type t = O.t list
      module Eq = Eq_list{O.Eq}
      let compare x y = List.compare O.compare x y
    end
\end{lstlisting}

The basic idea is to represent inheritance by including a submodule of
the inherited type, along with an implicit functor to extract that
submodule. By wrapping the inherited components in a module they can be
aliased.

The two sides of the ``diamond'' are now \code{Eq_list(Eq_ord(Ord_int))}
or \\ \code{Eq_ord(Ord_list(Ord_int))}, both of which are aliases for
\code{Eq_list(Eq_int)} so there is no ambiguity.

\subsection{Constructor classes}
\label{section:constructor-classes}

Since OCaml's modules support type members which have type parameters, modular
implicits naturally support \textit{constructor
  classes}~\cite{DBLP:journals/jfp/Jones95} -- i.e. functions whose implicit
instances are indexed by parameterised type constructors.  For example, here
is a definition of a \code{Functor} module type, together with implicit
instances for the parameterised types \code{list} and \code{option}:
\begin{lstlisting}
    module type Functor = sig
      type +'a t
      val map : ('a -> 'b) -> 'a t -> 'b t
    end

    let map {F: Functor} (f : 'a -> 'b) (c : 'a F.t) = F.map f c

    implicit module Functor_list = struct
      type 'a t = 'a list
      let map = List.map
    end

    implicit module Functor_option = struct
      type 'a t = 'a option
      let map f = function
         None -> None
       | Some x -> Some (F x)
    end
\end{lstlisting}
The choice to translate implicits into first-class functors makes
elaboration for implicit modules with parameterised types
straightforward.  Here is the elaborated code for \code{map}:
\begin{lstlisting}
    let map =
      (module functor (F: Functor) -> struct
         let value (f : 'a -> 'b) (c : 'a F.t) = F.map f c
       end)
\end{lstlisting}

\subsection{Multi-parameter type classes}
\label{section:overloading-on-multiple-types}

Most of the examples we have seen so far involve resolution of
implicit modules with a single type member.  However, nothing in the
design of modular implicits restricts resolution to a single type.
The module signature inclusion relation on which resolution is based
supports modules with an arbitrary number of type members (and indeed,
with many other components, such as modules and module types).

Here is an example illustrating overloading with multiple types.  The
\code{Widen} signature includes two type members, \code{slim} and \code{wide},
and a coercion function \code{widen} for converting from the former to the
latter.  The two implicit modules, \code{Widen_int_float} and
\code{Widen_opt}, respectively implement conversion from a \code{ints} to
\code{floats}, and lifting of widening to options.  The final line illustrates
the instantiation of a widening function from \code{int option} to \code{float option},
based on the three implicit modules.
\begin{lstlisting}
    module type Widen = sig
      type slim
      type wide
      val widen : slim -> wide
    end

    let widen {C:Widen} (v: C.slim) : C.wide = C.widen v

    implicit module Widen_int_float = struct
      type slim = int
      type wide = float
      let widen = Pervasives.float
    end

    implicit module Widen_opt{A: Widen} = struct
      type slim = A.slim option
      type wide = A.wide option
      let widen = function
         None -> None
       | Some v -> Some (A.widen v)
    end

    let v : float option = widen (Some 3)
\end{lstlisting}
In order to find a suitable implicit argument \code{C} for the call to
\code{widen} on the last line, the resolution procedure first generates
fresh types variables for \code{C.slim} and \code{C.wide}
\begin{lstlisting}
     C.slim = 'a
     C.wide = 'b
\end{lstlisting}
and replaces the corresponding names in the type of the variable \code{widen}:
\begin{lstlisting}
   widen : 'a -> 'b
\end{lstlisting}
Unifying this last type with the type supplied by the context
(i.e. the type of the argument and the ascribed result type) reveals that
\code{'a} should be equal to \code{int option} and \code{'b} should be equal
to \code{float option}.  The search for a suitable argument must therefore
find a module of type \code{Widen} with the following constraints:
\begin{lstlisting}
     C.slim = int option
     C.wide = float option
\end{lstlisting}
The implicit functor \code{Widen_option} is suitable if a modules \code{A} can
be found such that A has type \code{Widen} with the constraints
\begin{lstlisting}
    C.slim = int
    C.wide = float
\end{lstlisting}
The implicit module \code{Widen_int_float} satisfies these constraints, and
the search is complete.

The instantiated call shows the implicit module argument constructed
by the resolution procedure:

\begin{lstlisting}
    let v : float option =
      widen {Widen_option(Widen_int_float)} (Some 3)
\end{lstlisting}

\subsection{Associated types}
\label{features:associated-types}

Since OCaml modules can contain abstract types, searches can be existentially
quantified.  For example, we can ask for a type which can be shown
\begin{lstlisting}
    Show
\end{lstlisting}
rather than how to show a specific type
\begin{lstlisting}
    Show with type t = int
\end{lstlisting}

The combination of signatures with multiple type members and support
for existential searches gives us similar features to Haskell's
associated types~\cite{DBLP:conf/popl/ChakravartyKJM05}. We can search
for a module based on a subset of the types it contains and the search
will fill-in the remaining types for us. For example, here is a module
type \code{Array} for arrays with a type \code{t} of arrays and a type
\code{elem} of array elements, together with a function \code{create}
for creating arrays:
\begin{lstlisting}
    module type Array = sig
      type t
      type elem
      [...]
    end

    val create : {A : Array} -> int -> A.elem -> A.t
\end{lstlisting}
The \code{create} function can be used without specifying the array
type being created:
\begin{lstlisting}
    let x = create 5 true
\end{lstlisting}
This will search for an implicit \code{Array with type elem =
  bool}. When one is found \code{x} will correctly be given the
associated \code{t} type. This allows different array types to be used
for different element types. For example, arrays of bools could be
implemented as bit vectors, and arrays of ints implemented using regular
OCaml arrays by placing the following declarations in scope:
\begin{lstlisting}
    implicit module Bool_array = Bit_vector

    implicit module Int_array = Array(Int)
\end{lstlisting}

\subsection{Backtracking}
\label{resolution:backtracking}

Haskell's type class system ignores instance constraints when
determining whether two instances are ambiguous.  For example, the following
two instance constraints are always considered ambiguous:
\begin{lstlisting}[style=haskell]
    instance Floating n => Complex (Complex_cartesian n)
    instance Integral n => Complex (Complex_cartesian n)
\end{lstlisting}

In contrast, our system only considers those implicit functors for which
suitable arguments are in scope as candidates for instantiation.  For
example, the following two implicit functors are not inherently
ambiguous:
\begin{lstlisting}
    implicit module Complex_cartsian_floating {N: Floating}
      : Complex with type t = N.t complex_cartesian
    implicit module Complex_cartsian_integral {N: Integral}
      : Complex with type t = N.t complex_cartesian
\end{lstlisting}

The \code{Complex_cartesian_floating} and \code{Complex_cartesian_integral}
modules only give rise to ambiguity if constraint generation
(Section~\ref{implicit-application:generating}) determines that the type
\code{t} of the \code{Complex} signature should be instantiated to \code{s
  complex_cartesian} where there are instances of both \code{Floating} and
\code{Integral} in scope for \code{s}:
\begin{lstlisting}
    implicit module Floating_s : Floating with type t = s
    implicit module Integral_s : Integral with type t = s
\end{lstlisting}

Taking functor arguments into account during resolution is a form of
backtracking. The resolution procedure considers both
\code{Complex_cartesian_integral} and \code{Complex_cartesian_floating}
as candidates for instantiation and attempts to find suitable arguments
for both.  The resolution is only ambiguous if both implicit functors
can be applied to give implicit modules of the appropriate type.

\subsection{Local instances}
\label{section:local-instances}

The \code{let implicit} construct described in
Section~\ref{section:new-syntax} makes it possible to define implicit modules
whose scope is limited to a particular expression.  The following example
illustrates how these local implicit modules can be used to select alternative
behaviours when calling overloaded functions.

Here is a signature \code{Ord}, for types which support comparison:
\begin{lstlisting}
    module type Ord = sig
      type t
      val cmp : t -> t -> int
    end
\end{lstlisting}
The \code{Ord} signature makes a suitable type for the implicit argument of a
\code{sort} function:
\begin{lstlisting}
    val sort : {O: Ord} -> O.t list -> O.t list
\end{lstlisting}
Each call to \code{sort} constructs a suitable value for \code{Ord} from the
implicit modules and functors in scope. Two possible orderings for \code{int}
are:
\begin{lstlisting}
    module Ord_int = struct
      type t = int
      let cmp l r = Pervasives.compare l r
    end

    module Ord_int_rev = struct
     type t = int
      let cmp l r = Pervasives.compare r l
    end
\end{lstlisting}
Either ordering can be used with \code{sort} by passing the argument
explicitly:
\begin{lstlisting}
    sort {Ord_int} items
\end{lstlisting}
or
\begin{lstlisting}
    sort {Ord_int_rev} items
\end{lstlisting}

Explicitly passing implicit arguments bypasses the resolution mechanism
altogether.  It is occasionally useful to combine overriding of implicit
modules for particular types with automatic resolution for other types.  For
example, if the following implicit module definition is in scope then
\code{sort} can be used to sort lists of pairs of integers:
\begin{lstlisting}
    implicit module Ord_pair {A: Ord} {B: Ord} = struct
      type t = A.t * B.t
      let cmp (x1, x2) (y1, y2) =
        let c = A.cmp x1 y1 in
        if c <> 0 then c else B.cmp x2 y2
    end
\end{lstlisting}
Suppose that we want to use \code{Ord_pair} together with both the
regular and reversed integer comparisons to sort a list of pairs.  One
approach is to construct and pass entire implicit arguments
explicitly:
\begin{lstlisting}
    sort {Ord_pair(Int_ord_rev)(Int_ord_rev)} items
\end{lstlisting}
Alternatively (and equivalently), local implicit module bindings for
\code{Ord} and \code{Ord_int_rev} make it possible to override the
behaviour at ints while using the automatic resolution behaviour to
locate and use the \code{Ord_pair} functor:
\begin{lstlisting}
    let sort_both_ways (items : (int * int) list) =
      let ord =
        let implicit module Ord = Ord_int in
          sort items
      in
      let rev =
        let implicit module Ord = Ord_int_rev in
          sort items
      in
        ord, rev
\end{lstlisting}

In Haskell, which lacks both local instances and a way of explicitly
instantiating type class dictionary arguments, neither option is available,
and programmers are advised to define library functions in pairs, with one
function (such as \code{sort}) that uses type classes to instantiate arguments
automatically, and one function (such as \code{sortBy}) that accepts a regular
argument in place of a dictionary:
\begin{lstlisting}[style=haskell]
    sort :: Ord a => [a] -> [a]
    sortBy :: (a -> a -> Ordering) -> [a] -> [a]
\end{lstlisting}

\subsection{Structural matching}
\label{section:structural-matching}

As Section~\ref{resolution:modules} explains, picking a suitable implicit
argument involves a module which matches a constrained signature.  In contrast
to Haskell's type classes, matching is therefore defined structurally (in
terms of the names and types of module components) rather than nominally (in
terms of the name of the signature).  Structural matching allows the caller of
an overloaded function to determine which part of a signature is required
rather than requiring the definer of a class to anticipate which overloaded
functions are most suitable for grouping together.

It is not difficult to find situations where structural matching is useful.
The following signature describes types which support basic arithmetic, with members
for zero and one, and for addition and subtraction:
\begin{lstlisting}
    module type Num = sig
      type t
      val zero : t
      val one : t
      val ( + ) : t -> t -> t
      val ( * ) : t -> t -> t
    end
\end{lstlisting}
The following implicit modules implement \code{Num} for the types \code{int}
and \code{float}, using functions from OCaml's standard library:
\begin{lstlisting}
    implicit module Num_int = struct
      type t = int
      let zero = 0
      let one = 1
      let ( + ) = Pervasives.( + )
      let ( * ) = Pervasives.( * )
    end

    implicit module Num_float = struct
      type t = float
      let zero = 0.0
      let one = 1.0
      let ( + ) = Pervasives.( +. )
      let ( * ) = Pervasives.( *. )
    end
\end{lstlisting}
The \code{Num} signature makes it possible to define a variety of arithmetic
functions.  However, in some cases \code{Num} offers more than necessary.  For
example, defining an overloaded function \code{sum} to compute the sum of a
list of values requires only \code{zero} and \code{+}, not \code{one} and
\code{*}.  Using \code{Num} as the implicit signature for \code{sum} would
make unnecessarily exclude types (such as strings) which have a notion of
addition but which do not support multiplication.

Defining more constrained signatures makes it possible to define more general
functions.  Here is a signature \code{Add} which includes only those elements
of \code{Num} involved in addition:
\begin{lstlisting}
    module type Add = sig
      type t
      val zero : t
      val ( + ) : t -> t -> t
    end
\end{lstlisting}
Using \code{Add} we can define a \code{sum} which works for any type that
has an implicit module with definitions of \code{zero} and \code{plus}:
\begin{lstlisting}
    let sum {A: Add} (l : A.t list) =
      List.fold_left A.( + ) A.zero l
\end{lstlisting}
The existing implicit modules \code{Num_int} and \code{Num_float} can be used
with \code{sum}, since they both match \code{Add}.  The following module,
\code{Add_string}, also matches \code{Add}, making it possible to use
\code{sum} either for summing a list of numbers or for concatenating a list of
strings:
\begin{lstlisting}
    implicit module Add_string = struct
      type t = string
      let zero = ""
      let ( + ) = Pervasives.( ^ ) (* concatenation *)
    end
\end{lstlisting}

In other cases it may be necessary to use some other part of the \code{Num}
interface.  The following function computes an inner product for any type with
an implicit module that matches \code{Num}:
\begin{lstlisting}
    let dot {N: Num} (l1 : N.t list) (l2 : N.t list) =
      sum (List.map2 N.( * ) l1 l2)
\end{lstlisting}
This time it would not be sufficient to use \code{Add} for the type of the
implicit argument, since \code{dot} uses both multiplication and addition.
However, \code{Add} still has a role to play: the implicit argument of
\code{sum} uses the implicit argument \code{N} with type \code{Add}.  Since
the \code{Num} signature is a subtype of \code{Add} according to the rules of
OCaml's module system, the argument can be passed through directly to
\code{sum}.  Here is the elaboration of \code{dot}, showing how the \code{sum}
functor being unpacked, bound to \code{F}, then applied to the implicit
argument \code{N}:
\begin{lstlisting}
    let dot =
      (module functor (N: Num) -> struct
         let value (l1 : N.t list) (l2 : N.t list) =
           let module F = (val sum) in
           let module R = F(N) in
             R.value (List.map2 N.( * ) l1 l2)
       end)
\end{lstlisting}
An optimising compiler might lift the unpacking and application of \code{sum}
outside the body of the function, in order to avoid repeating the work each
time the list arguments are supplied.

Section~\ref{section:inheritance} illustrated that structural matching
is not an ideal encoding for full class inheritance hierarchies due to
the diamond problem. However, it can provide a more lightweight encoding
for simple forms of inheritance.

\section{Canonicity}
\label{section:canonicity}

In Haskell, a type class has at most one instance per type within a
program. For example, defining two instances of \code{Show} for the type
\code{Int} or for the type constructor \code{Maybe} is not permitted.
We call this property \emph{canonicity}.

Haskell relies on canonicity to maintain coherence, whereas canonicity
cannot be preserved by our system due to OCaml's support for modular
abstraction.

\subsection{Inference, coherence and canonicity}
\label{canonicity:inference}

A key distinction between type classes and implicits is that, with type
classes, constraints on a function's type can be inferred based on the
use of other constrained functions in the function's definitions. For
example, if a \code[haskell]{show_twice} function uses the
\code[haskell]{show} function:
\begin{lstlisting}[style=haskell]
    show_twice x = show x ++ show x
\end{lstlisting}
then Haskell will infer that \code[haskell]{show_twice} has type
\code[haskell]{Show a => a -> String}.

This inference raises issues for coherence in languages with type
classes. For example, suppose we have the following instance:
\begin{lstlisting}[style=haskell]
    instance Show a => Show [a] where
        show l = show_list l
\end{lstlisting}
and consider the function:
\begin{lstlisting}[style=haskell]
    show_as_list x = show [x]
\end{lstlisting}
There are two valid types which could be inferred for this function:
\begin{lstlisting}[style=haskell]
    show_as_list :: Show [a] => a -> String
\end{lstlisting}
or
\begin{lstlisting}[style=haskell]
    show_as_list :: Show a => a -> String
\end{lstlisting}
In the second case, the \code[haskell]{Show [a]} instance has been used
to reduce the constraint to \code[haskell]{Show a}.

The choice between these two types changes where the \code[haskell]{Show
  [a]} constraint is resolved. In the first case it will be resolved at
\emph{calls} to \code[haskell]{show_as_list}. In the second case it has
been resolved at the \emph{definition} of \code[haskell]{show_as_list}.

If type class instances are canonical then it does not matter where a
constraint is resolved, as there is only a single instance to which it
could be resolved. Thus, with canonicity, the inference of
\code[haskell]{show_as_list}'s type cannot affect the dynamic semantics
of the program, and coherence is preserved.

However, if type class instances are not canonical then where a
constraint is resolved can affect which instance is chosen, which in
turn changes the dynamic semantics of the program. Thus, without
canonicity, the inference of \code[haskell]{show_as_list}'s type can
affect the dynamic semantics of the program, breaking coherence.

\subsection{Canonicity and abstraction}
\label{canonicity:abstraction}

It would not be possible to preserve canonicity in OCaml because type
aliases can be made abstract. Consider the following example:
\begin{lstlisting}
    module F (X : Show) = struct
      implicit module S = X
    end

    implicit module Show_int = struct
      type t = int
      let show = string_of_int
    end

    module M = struct
      type t = int
      let show _ = "An int"
    end

    module N = F(M)
\end{lstlisting}
The functor \code{F} defines an implicit \code{Show} module for the
abstract type \code{X.t}, whilst the implicit module \code{Show_int} is
for the type \code{int}. However, \code{F} is later applied to a module
where \code{t} is an alias for \code{int}. This violates canonicity but
this violation is hidden by abstraction.

Whilst it may seem that such cases can be detected by peering through
abstractions, this is not possible in general and defeats the entire
purpose of abstraction. Fundamentally, canonicity is not a modular
property and cannot be respected by a language with full support for
modular abstraction.

\subsection{Canonicity as a feature}
\label{canonicity:feature}

Besides maintaining coherence, canonicity is sometimes a useful feature
in itself.  The canonical example for the usefulness of canonicity is
the \code{union} function for sets in Haskell. The \code[haskell]{Ord}
type class defines an ordering for a type\footnote{Some details of
  \code[haskell]{Ord} are omitted for simplicity}:
\begin{lstlisting}[style=haskell]
    class Ord a where
      (<=) :: a -> a -> Bool
\end{lstlisting}
This ordering is used to create sets implemented as binary trees:
\begin{lstlisting}[style=haskell]
    data Set a
    empty :: Set a
    insert :: Ord a => a -> Set a -> Set a
    delete :: Ord a => a -> Set a -> Set a
\end{lstlisting}
The \code[haskell]{union} function computes the union of two sets:
\begin{lstlisting}
    union :: Ord a => Set a -> Set a -> Set a
\end{lstlisting}
Efficiently implementing this union requires both sets to have been created
using the same ordering. This property is ensured by canonicity, since there
is only one instance of \code[haskell]{Ord a} for each \code[haskell]{a}, and
all sets of type \code[haskell]{Set a} must have been created using it.

\subsection{An alternative to canonicity as a feature}
\label{canonicity:alternative}

In terms of modular implicits, Haskell's \code[haskell]{union} function would
have type:
\begin{lstlisting}
    val union: {O : Ord} -> O.t set -> O.t set -> O.t set
\end{lstlisting}
but without canonicity it is not safe to give \code{union} this type
since there is no guarantee that all sets of a given type were created
using the same ordering.

The issue is that the \code{set} type is only parametrised by the type
of its elements, when it should really be also parametrised by the ordering
used to create it. Traditionally, this problem is solved in OCaml by
using applicative functors:
\begin{lstlisting}
    module Set (O : Ord) : sig
      type elt
      type t
      val empty : t
      val add : elt -> t -> t
      val remove : elt -> t -> t
      val union : t -> t -> t
      [...]
    end
\end{lstlisting}
When applied to an \code{Ord} argument \code{O}, the \code{Set} functor
produces a module containing the following functions:
\begin{lstlisting}
    val empty : Set(O).t
    val add : elt -> Set(O).t -> Set(O).t
    val remove : elt -> Set(O).t -> Set(O).t
    val union : Set(O).t -> Set(O).t -> Set(O).t
\end{lstlisting}

The same approach transfers to modular implicits, giving our polymorphic
set operations the following types:
\begin{lstlisting}
    val empty : {O : Ord} -> Set(O).t
    val add : {O : Ord} -> O.t -> Set(O).t -> Set(O).t
    val remove : {O : Ord} -> O.t -> Set(O).t -> Set(O).t
    val union : {O : Ord} -> Set(O).t ->
                  Set(O).t -> Set(O).t
\end{lstlisting}
The type for sets is now \code{Set(O).t} which is parametrised by the
ordering module \code{O}, ensuring that \code{union} is only applied to
sets created using the same ordering.

\section{Order independence and compositionality}
\label{section:subtleties}

Two properties enjoyed by traditional ML type systems are \emph{order
  independence} and \emph{compositionality}. This section describes how
modular implicits affect these properties.

\subsection{Order independence}
\label{implicit-application:dependence}

Type inference is \emph{order independent} when the order in which
expressions are type-checked does not affect whether type inference
succeeds. Traditional ML type inference is order independent, however
some of OCaml's advanced features, including first-class functors, cause
order dependence.

As described in Section~\ref{resolution}, type checking implicit applications
has two aspects:
\begin{enumerate}
\item Inferring the types which constrain the implicit argument
\item Resolving the implicit argument using the modules and functors in
  the implicit scope.
\end{enumerate}
These two aspects are interdependent: the order in which they are
performed affects whether type inference succeeds.

\subsubsection{Resolution depends on types}
\label{implicit-application:dependence:resolution}

Consider the implicit application from line 24 of our \code{Show}
example (Figure~\ref{fig:show}):
\begin{lstlisting}
    show 5
\end{lstlisting}
Resolving the implicit argument \code{S} requires first generating the
constraint \code{S.t = int}. Without this constraint the argument would be
ambiguous -- it could be \code{Show_int}, \code{Show_float}, \\
\code{Show_list(Show_float)}, etc. This constraint can only come from
type-checking the non-implicit argument \code{5}.

This demonstrates that resolution depends on type inference, and so some type
inference must be done before implicit arguments are resolved.

\subsubsection{Types depend on resolution}
\label{implicit-application:dependence:types}

Given that resolution depends on type inference, we might be tempted to
perform resolution in a second pass of the program, after all type inference
has finished.  However, although it is not immediately obvious, types also
depend on resolution.

Consider the following code:
\begin{lstlisting}
    module type Sqrtable = sig
      type t
      val sqrt : t -> t
    end

    let sqrt {S : Sqrtable} x = S.sqrt x

    implicit module Sqrt_float = struct
      type t = float
      let sqrt x = sqrt_float x
    end

    let sqrt_twice x = sqrt (sqrt x)
\end{lstlisting}
The \code{sqrt_twice} function contains two calls to \code{sqrt}, which
has an implicit argument of module type \code{Sqrtable}. There are no
constraints on these implicit parameters as \code{x} has an unknown
type; however, there is only one \code{Sqrtable} module in scope so the
resolution is still unambiguous. By resolving \code{S} to
\code{Sqrt_float} we learn that \code{x} in fact has type \code{float}.

This demonstrates that types depend on resolution, and so resolution
must be done before some type inference. In particular, it is important
that resolution is performed before generalisation is attempted on any
types which depend on resolution because type variables cannot be
unified after they have been generalised.

\subsubsection{Resolution depends on resolution}
\label{implicit-application:dependence:circular}

Since resolution depends on types, and types can depend on resolution,
it follows that one argument's resolution can depend on another
argument's resolution.

Following on from the previous example, consider the following code:
\begin{lstlisting}
    module type Summable = sig
      type t
      val sum : t -> t -> t
    end

    let double {S : Summable} x = S.sum x x

    implicit module Sum_int = struct
      type t = int
      let sum x y = x + y
    end

    implicit module Sum_float = struct
      type t = float
      let sum x y = x +. y
    end

    let sqrt_double x = sqrt (double x)
\end{lstlisting}
Here there are two implicit applications: one of \code{sqrt} and one of
\code{double}. As before, the arguments of these functions have no
constraints since \code{x}'s type is unknown. If the resolution of
\code{double}'s implicit argument is attempted without constraint it
will fail as ambiguous, since either \code{Sum_int} or \code{Sum_float}
could be used. However, if \code{sqrt}'s implicit argument is first
resolved to \code{Sqrt_float} then we learn that the return type of the
call to \code{double} is float. This allows \code{double}'s implicit
argument to unambiguously be resolved as \code{Sum_float}.

This demonstrates that resolutions can depend on other resolutions,
and so the order in which resolutions are attempted will affect which
programs will type-check successfully.

\subsubsection{Predictable inference}
\label{implicit-application:dependence:predictable}

In the presence of order dependence, inference can be kept predictable
by providing some declarative guarantees about the order of
type-checking, and disallowing programs whose type inference would only
succeed due to an ordering between operations which is not ensured by
these guarantees. This is the approach OCaml takes with its other
order-dependent features\footnote{OCaml emits a warning
  rather than out-right disallowing programs which depend on an
  unspecified ordering}.

Taking the same approach with modular implicits involves two
choices about the design:
\begin{enumerate}
\item When should implicit resolution happen relative to type inference?
\item In what order should implicit arguments be resolved?
\end{enumerate}

The dependence of resolution on type inference is much stronger than the
dependence of type inference on resolution: delaying type inference
until after resolution would lead to most argument resolutions being
ambiguous.

In order to perform as much inference as possible before attempting
resolution, resolution is delayed until the point of
generalisation. Technically, resolution could be delayed until a
generalisation is reached which directly depends on a type involved in
that resolution. However, we take a more predictable approach and
resolve all the implicit arguments in an expression whenever the result
of that expression is generalised.

In practice, this means that implicit arguments are resolved at the
nearest enclosing let binding. For example, in this code:
\begin{lstlisting}
    let f g x =
      let z = [g (show 5) (show 4.5); x] in
        g x :: z
\end{lstlisting}
the implicit arguments of both calls to \code{show} will be resolved
after the entire expression
\begin{lstlisting}
    [g (show 5) (show 4.5); x]
\end{lstlisting}
has been type-checked, but before the expression
\begin{lstlisting}
    g x :: z
\end{lstlisting}
has been type-checked.

Our implementation of modular implicits makes very few guarantees about the
order of resolution of implicit arguments within a given expression. It is
guaranteed that implicit arguments of the same function will be resolved
left-to-right, and that implicit arguments to a function will be resolved
before any implicit arguments within other arguments to that function.

These guarantees mean that the example of dependent resolutions:
\begin{lstlisting}
    let sqrt_double x = sqrt (double x)
\end{lstlisting}
will resolve without ambiguity, but that the similar expression:
\begin{lstlisting}
    let double_sqrt x = double (sqrt x)
\end{lstlisting}
will result in an ambiguous resolution error. This can be remedied either
by adding a type annotation:
\begin{lstlisting}
    let double_sqrt x : float = double (sqrt x)
\end{lstlisting}
or by lifting the argument into its own \code{let} expression to force its
resolution:
\begin{lstlisting}
    let double_sqrt x =
      let s = sqrt x in
        double s
\end{lstlisting}

Another possibility would be to try each implicit argument being
resolved in turn until an unambiguous one is found. Resolving that
argument might produce more typing information allowing further
arguments to be resolved unambiguously. This approach is analogous to
using a breadth-first resolution strategy in logic programming, rather
than a depth-first strategy: it improves the completeness of the search
-- and so improves the predictability of inference -- but is potentially
less efficient in practice. Comparing this approach with the one used in
our existing implementation is left for future work.

\subsection{Compositionality}
\label{resolution:compositionality}

Compositionality refers to the ability to combine two independent
well-typed program fragments to produce a program fragment that is also
well typed. In OCaml, this property holds of top-level definitions up to
renaming of identifiers.

Requiring that implicit arguments be unambiguous means that renaming of
identifiers is no longer sufficient to guarantee two sets of top-level
definitions can be composed. For example,
\begin{lstlisting}
    implicit module Show_int1 = struct
      type t = int
      let show x = "Show_int1: " ^ (int_of_string x)
    end

    let x = show 5
\end{lstlisting}
and
\begin{lstlisting}
    implicit module Show_int2 = struct
      type t = int
      let show x = "Show_int2: " ^ (int_of_string x)
    end

    let y = show 6
\end{lstlisting}
cannot be safely combined because the call to \code{show} in the
definition of \code{y} would become ambiguous. In order to ensure that
two sets of definitions can safely compose they must not contain
overlapping implicit module declarations.

However, whilst compositionality of top-level definitions is lost,
compositionality of modules is maintained. Any two well-typed module
definitions can be combined to produce a well-typed program. This is an
important property, as it allows support for separate compilation
without the possibility of errors at link time.

\section{Implementation}
\label{implementation}

We have created a prototype implementation of our proposal based on
OCaml 4.02.0, which can be installed through the OPAM package manager:

\begin{lstlisting}
    opam switch 4.02.0+modular-implicits
\end{lstlisting}

Although it is not yet in a production ready state, the prototype has
allowed us to experiment with the design and to construct examples
like those in Section~\ref{section:modular-implicits-by-example}. We
have also used the prototype to build a port of Haskell's Scrap Your
Boilerplate~\cite{Lammel:2003:SYB:640136.604179} library, which
involves aroud 600 lines of code, and exercises many of the features
and programming patterns described in this paper, including
inheritance, higher-order polymorphism, implicit functors and
constructor classes.  As the prototype becomes more stable we hope to
use it to explore the viability of modular implicits at greater scale.

One key concern when implementing modular implicits is the efficiency
of the resolution procedure. Whilst the changes to OCaml's type
inference required for modular implicits are small and should not
affect its efficiency, the addition of a resolution for every use of
functions with implicit parameters could potentially have a dramatic
effect on performance.

Our prototype implementation takes a very naive approach to
resolution, keeping a list of the implicit modules and functors in
scope, and checking each of them as a potential solution using OCaml's
existing procedure for checking module inclusion.

The performance of resolution could be improved in the following ways:
\begin{description}
\item[Memoization] Resolutions can be memoized so that repeated uses
  of the same functions with implicit arguments do not cause repeated
  full resolutions. Even if new implicit modules are added to the
  environment it is possible to partially reuse the results of
  previous resolutions since module expressions which do not involve
  the new modules do not need to be reconsidered.
\item[Indexing] A mapping can be maintained between module types and the
  implicit modules which could be used to resolve them to avoid
  searching over the whole list of implicit modules in scope. In
  particular, indexing based on the names of the members of the module
  type is simple to implement and should quickly reduce the number of
  implicit modules that need to be considered for a particular
  resolution.
\item[Fail-fast module inclusion] Checking module inclusion in OCaml
  is an expensive operation. However, during resolution most inclusion
  checks are expected to fail. Using an implementation of module
  inclusion checking which is optimised for the failing case would
  make it possible to quickly eliminate most potential solutions to a
  resolution.
\end{description}

These techniques aim to reduce the majority of resolutions to a few
table lookups, which should allow modular implicits to scale
effectively to large code bases with many implicit declarations and
implicit arguments. However, we leave the implementation and full
evaluation of these techniques to future work.

\section{Related work}
\label{related-work}

There is a large literature on systematic approaches to ad-hoc
polymorphism, Kaes~\cite{DBLP:conf/esop/Kaes88} being perhaps the
earliest example. We restrict our attention here to a representative
sample.

\subsection{Type classes}
\label{related-work:type-classes}

Haskell type classes~\cite{DBLP:conf/popl/WadlerB89} are the classic
formalised method for ad-hoc polymorphism. They have been replicated in
a number of other programming languages (e.g. Agda's instance
arguments~\cite{DBLP:conf/icfp/DevrieseP11}, Rust's traits~\cite{rust}).

The key difference between approaches based on type class and approaches
based on implicits is that type class constraints can be inferred,
whilst implicit parameters must be defined explicitly. Haskell
maintains coherence, in the presence of such inference, by ensuring that
type class instances are canonical.

Canonicity is not possible in a language which supports modular abstraction
(such as OCaml), and so type classes are not always a viable
choice. Canonicity is also not always desirable: the restriction to a single
instance per type is not compositional and can force users to create
additional types to work around it.  Consequently, some proposals for
extensions to type classes involve relinquishing canonicity in order to
support desirable features such as local
instances~\cite{dijkstra_et_al_scoped_instances_07}.

The decision to infer constraints also influences other design
choices.  For example, whereas modular implicits instantiate implicit
arguments only at function application sites, the designers of type
classes take the dual approach of only generalizing constrained type
variables at function abstraction
sites~\cite[Section~4.5.5]{Haskell98}.  Both restrictions have the
motivation of avoiding unexpected work -- in Haskell, adding
constraints to non-function bindings can cause a loss of sharing,
whereas in OCaml, inserting implicit arguments at sites other than
function calls could cause side effects to take place in the evaluation
of apparently effect-free code.

Modular implicits offer a number of other advantages over type classes,
including support for backtracking during parameter resolution, allowing for
more precise detection of ambiguity, and resolution based on any type defined
within the module rather than on a single specific type.  However, there are
also some features of type classes that our proposal does not support, such as
the ability to instantiate an instance variable with an open type expression;
in Haskell one can define the following instance, which makes it possible to
show values of type \code{T a} for any \code{a}:

\begin{lstlisting}[style=haskell]
    instance Show (T a)
\end{lstlisting}

\subsection{Implicits}
\label{related-work:implicits}

Scala implicits~\cite{DBLP:conf/oopsla/OliveiraMO10} are a major
inspiration for this work. They provide implicit parameters on
functions, which are selected from the scope of the call site based on
their type. In Scala these parameters have ordinary Scala types, whilst
we propose using module types. Scala's object system has many properties
in common with a module system, so advanced features such as associated
types are still possible despite Scala's implicits being based on
ordinary types.

Scala's implicits have a more complicated notion of scope than our
proposal. This seems to be aimed at fitting implicits into Scala's
object-oriented approach: for example allowing implicits to be searched
for in companion objects of the class of the implicit parameter. This
makes it more difficult to answer the question ``Where is the implicit
parameter coming from?'', in turn making it more difficult to reason
about code. Our proposal simply uses lexical scope when searching for an
implicit parameter.

Scala supports overlapping implicit instances. If an implicit parameter
is resolved to more than one definition, rather than give an ambiguity
error, a complex set of rules gives an ordering between definitions, and
a most specific definition will be selected. An ambiguity error is only
given if multiple definitions are considered equally specific. This can
be useful, but makes reasoning about implicit parameters more
difficult: to know which definition is selected you must know all the
definitions in the current scope. Our proposal always gives an ambiguity
error if multiple implicit modules are available.

In addition to implicit parameters, Scala also supports implicit
conversions. If a method is not available on an object's type the
implicit scope is searched for a function to convert the object to a
type on which the method is available. This feature greatly increases
the complexity of finding a method's definition, and is not supported in
our proposal.

Chambart et al.\ have proposed~\cite{conf/oud/ChambartH12} adding
support for implicits to OCaml using core OCaml types for implicit
parameters. Our proposal instead uses module types for implicit
parameters.  This allows our system to support more advanced features
including associated types and higher kinds. The module system also
seems a more natural fit for ad-hoc polymorphism due to its direct
support for signatures.

The implicit calculus~\cite{DBLP:conf/pldi/OliveiraSCLY12} provides a
minimal and general calculus of implicits which could serve as a basis
for formalising many aspects of our proposal.

Coq's type classes~\cite{DBLP:conf/tphol/SozeauO08} are similar to
implicits. They provide implicit dependent record parameters selected
based on their type.

\subsection{Canonical structures}
\label{related-work:canonical-structures}

In addition to type classes, Coq also supports a mechanism for ad-hoc
polymorphism called \emph{canonical
  structures}~\cite{canonical-structures}. Type classes and implicits provide
a mechanism to resolve a value based on type information. Coq, being
dependently typed, already uses unification to resolve values from type
information, so canonical structures support ad-hoc polymorphism by providing
additional ad-hoc rules that are applied during unification.

Like implicits, canonical structures do not require canonicity, and do
not operate on a single specific type: ad-hoc unification rules are
created for every type or term defined in the structure. Canonical
structures also support backtracking of their search due to the
backtracking built into Coq's unification.

\subsection{Concepts}
\label{related-work:concepts}

Gregor et al.~\cite{DBLP:conf/oopsla/GregorJSSRL06} describe
\emph{concepts}, a system for ad-hoc polymorphism in C++\footnote{This
  should not be confused with more recent ``concepts lite'' proposal,
  due for inclusion in the next C++ standard}.

C++ has traditionally used simple overloading to support ad-hoc
polymorphism restricted to mono\-morphic uses. C++ also supports
parametric polymorphism through templates. However, overloading within
templates is re-resolved after template instantiation. This means that
the combination of overloading and templates provides full ad-hoc
polymorphism. Delaying a significant part of type checking until
template instantiation increases compilation times and makes error message
more difficult to understand.

Concepts provide a disciplined mechanism for full ad-hoc polymorphism
through an approach similar to type classes and implicits. Like type
classes, a new kind of type is used to constrain parametric type
variables. New concepts are defined using a \code[cpp]{concept}
construct. Classes with the required members of a concept automatically
have an instance for that concept, and further instances can be defined
using the \code[cpp]{concept_map} construct. Like implicits, concepts
cannot be inferred and are not canonical.

Concepts allow overlapping instances, using C++'s complex overloading
rules to resolve ambiguities. Concept maps can override the default
instance for a type.  These features can be useful, but make reasoning
about implicit parameters more difficult. Our proposal requires all
implicit modules to be explicit and always gives an ambiguity error if
multiple matching implicit modules are available.

F\#'s static constraints~\cite{syme2005f} are similar to concepts without
support for concept maps.

\subsection{Modular type classes}
\label{related-work:modular-type-classes}

Dreyer et al.~\cite{DBLP:conf/popl/DreyerHCK07} describe \emph{modular
  type classes}, a type class system which uses ML module types as type
classes and ML modules as type class instances. This system sticks
closely to the design of Haskell type classes.  In particular it infers
type class constraints, and gives ambiguity errors at the point when
modules are made implicit.

In order to maintain coherence in the presence of inferred constraints
and without canonicity, the system includes a number of undesirable
restrictions:
\begin{itemize}
\item Modules may only be made implicit at the top-level; they cannot be
  introduced within a module or a value definition.
\item Only module definitions are permitted at the top-level; all value
  definitions must be contained within a sub-module.
\item All top-level module definitions must have an explicit signature.
\end{itemize}
These restrictions essentially split the language into an outer layer
that consists only of module definitions and an inner layer within each
module definition. Within the inner layer instances are canonical and
constraints are inferred. In the outer layer instances are not
canonical and all types must be given explicitly; there is no type
inference.

In order to give ambiguity errors at the point where modules are made
implicit, one further restriction is required: all implicit modules must
define a type named \code{t} and resolution is always done based on this
type.

By basing our design on implicits rather than type classes we avoid such
restrictions. Our proposal also includes higher-rank implicit
parameters, higher-kinded implicit parameters and resolution based on
multiple types. These are not included in the design of modular type
classes.

Wehr et al.~\cite{DBLP:conf/aplas/WehrC08} give a comparison and
translation between modules and type classes. This translation does not
consider the implicit aspect of type classes, but does illustrate the
relationship between type class features (e.g. associated types) and
module features (e.g. abstract types).

\section{Future work}
\label{future-work}

This paper gives only an informal description of the type system and
resolution procedure. Giving a formal description is left as future
work.

The implementation of our proposal described in
Section~\ref{implementation} is only a prototype. Further work is needed
to bring this prototype up to production quality.

The two aspects of our proposal related to completeness of type
inference:
\begin{enumerate}
\item The interdependence of type inference and resolution
\item The restrictions on resolution to avoid non-termination
\end{enumerate}
are inevitably compromises between maximising inference and maximising
predictability. How to strike the best balance between these two goals
is an open question. More work is needed to evaluate how predictable
users find the various possible approaches in practice.

The syntax used for implicit functors is suggestive of an extension to
our proposal: functors with implicit arguments. In our proposal,
arguments to functors are only resolved implicitly during resolution for
other implicit arguments. Supporting such resolution more generally
would be an interesting direction to explore as it would introduce
ad-hoc polymorphism into the module language.

Further work is also needed to answer more practical questions: How well
do modular implicits scale to large code bases? How best to design
libraries using implicits? How efficient is implicit resolution on
real world code bases?

\subsubsection*{Acknowledgements}
We thank Stephen Dolan and Anil Madhavapeddy for helpful discussions,
Andrew Kennedy for bringing Canonical Structures to our attention, and
Oleg Kiselyov and the anonymous reviewers for their insightful
comments.

\bibliographystyle{eptcs}
\bibliography{bibliography.bib}

\end{document}